\documentclass[11pt]{article}
\usepackage{booktabs}
\usepackage[preprint]{acl}

\usepackage{times}
\usepackage{latexsym}
\usepackage{amsmath}
\usepackage{longtable}
\usepackage[most]{tcolorbox}
\usepackage{cleveref}
\usepackage{multirow}
\usepackage{tabularx}
\usepackage{array}
\usepackage{nccmath}
\usepackage{caption}
\usepackage{amssymb}
\usepackage{colortbl}
\usepackage{graphicx}
\tcbuselibrary{breakable}
\usepackage{wrapfig}
\usepackage{xcolor}
\usepackage{enumitem}
\newcommand{\ie}{\textit{i.e.}}
\newcommand{\eg}{\textit{e.g.}}

\usepackage[T1]{fontenc}

\usepackage[utf8]{inputenc}

\usepackage{microtype}

\usepackage{inconsolata}

\usepackage{graphicx}

\title{SumRank: Aligning Summarization Models for Long-Document \\ Listwise Reranking}

\author{
  \textbf{Jincheng Feng}, 
  \textbf{Wenhan Liu}, 
  \textbf{Zhicheng Dou\thanks{Corresponding authors.}},
  \\
  Gaoling School of Artificial Intelligence, Renmin University of China
\\
  \texttt{fengjincheng03@gmail.com, dou@ruc.edu.cn}
}

\begin{document}
\maketitle
\begin{abstract}
Large Language Models (LLMs) have demonstrated superior performance in listwise passage reranking. However, directly applying them to rank long-form documents introduces both effectiveness and efficiency issues due to the substantially increased context length. To address this challenge, we propose a per-document summarization model, SumRank, aligned with downstream listwise reranking, to compress long-form documents into concise rank-aligned summaries before the final listwise reranking stage. To develop our summarization model SumRank, we introduce a three-stage training pipeline consisting of cold-start Supervised Fine-Tuning (SFT), specialized Reinforcement Learning (RL) data construction, and rank-driven alignment via RL. This paradigm aligns the SumRank with downstream ranking objectives to preserve crucial relevance signals. We conduct extensive experiments on both in-domain (ID) benchmark datasets from the TREC Deep Learning tracks and out-of-domain (OOD) datasets from the BEIR benchmark. Results show that our lightweight SumRank model achieves state-of-the-art (SOTA) ranking performance while significantly improving efficiency by reducing both summarization overhead and reranking complexity.

\end{abstract}

\section{Introduction}

\begin{figure}[t] 
    \centering
    \includegraphics[trim={0.0cm 0.0cm 0.0cm 0.0cm}, clip, width=\columnwidth]{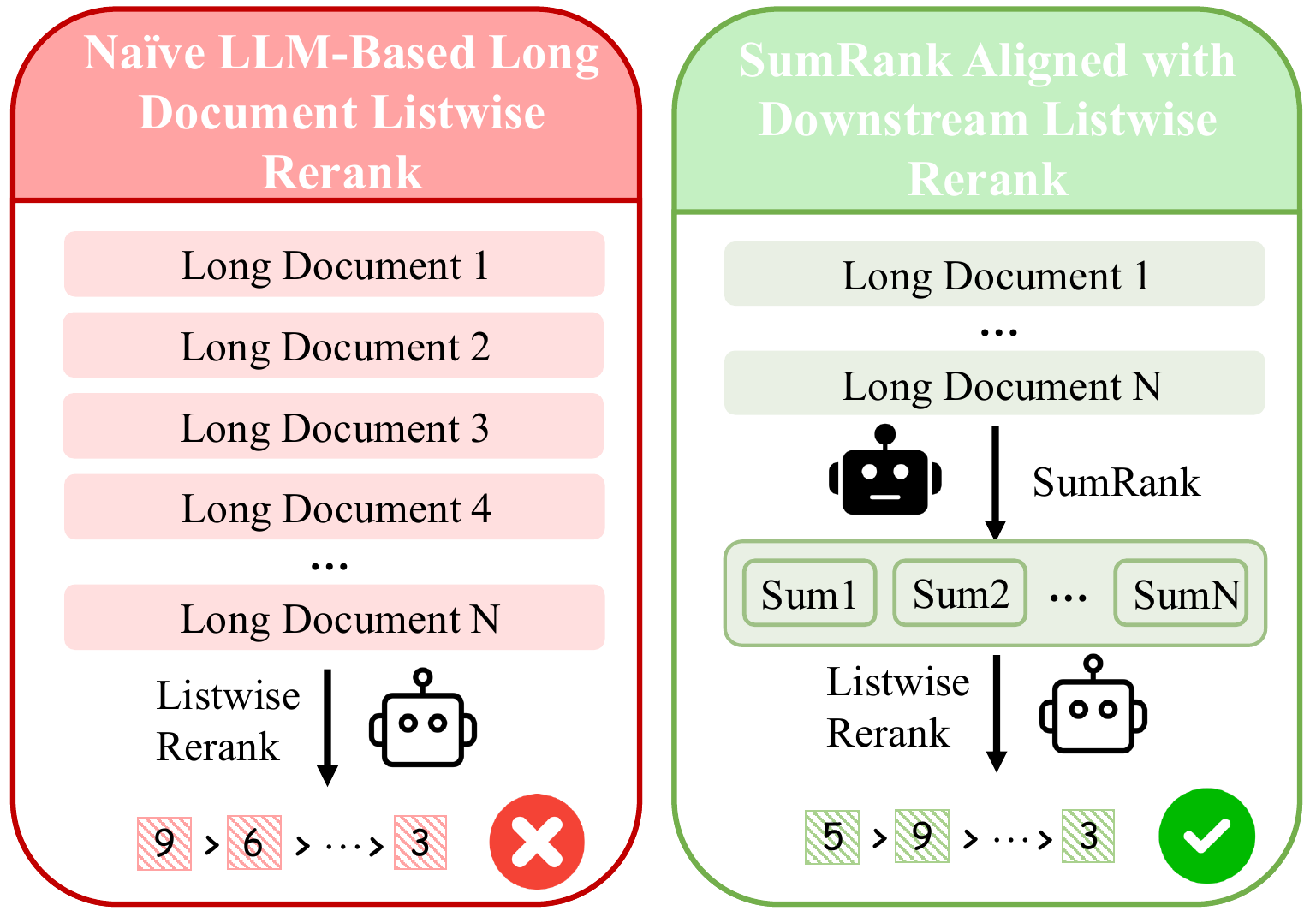}
    \caption{The left part shows the method of naive LLM-based long document listwise reranking. The right part shows our proposed SumRank, which is better than the naive listwise method both in effectiveness and efficiency. $\text{Sum}_1, \text{Sum}_2, \dots, \text{Sum}_N$ denote the summaries generated by SumRank.}
    \label{fig:intro}
\end{figure}

Document ranking plays a central role in Information Retrieval (IR) by reranking initially retrieved documents to improve the quality of search results. Recently, Large Language Models (LLMs) have demonstrated strong capabilities in zero-shot passage ranking~\cite{DBLP:journals/corr/abs-2308-07107}. Among various LLM-based ranking paradigms, listwise ranking has emerged as particularly effective ~\cite{rankgpt, reasonrank}. Instead of evaluating documents independently, listwise approaches consider an entire candidate list jointly, enabling the model to capture global relevance patterns across documents. Compared with pointwise~\cite{rankt5} and pairwise ranking~\cite{QinJHZWYSLLMWB24}, listwise ranking leads to more accurate ranking decisions and has achieved state-of-the-art (SOTA) performance on many IR benchmarks~\cite{dl19,dl20,beir,bright}.


Despite this success, these studies~\cite{DBLP:conf/naacl/ZhuangQHWYWB24,rankgpt,rankt5, fullrank, reasonrank, rubricrank} primarily focus on reranking \textit{short-passage style documents}, where each candidate unit contains only a limited amount of text (\eg, around 100 words), while reranking \textit{long-form documents} has received little attention. In fact, long-document ranking is of greater practical importance in many real-world retrieval scenarios~\cite{DBLP:journals/corr/abs-2503-17407,longdocsurvey, lin2026beyond}. In modern search engines, retrieved results are typically full-length webpages (with thousands of words) rather than short passages, and relevant information may be distributed across different sections of a lengthy document. Accurately assessing the relevance of such long documents is therefore essential not only for traditional web search, but also for downstream applications that rely on long-document retrieval, such as deep search~\cite{webthinker,deepresearch,agentic-r, smartsearch, masksearch} and tool-use agent~\cite{toolstar, tool, etagent}, where long web pages are retrieved as supporting evidence.


However, directly applying listwise reranking to long documents introduces both effectiveness and efficiency issues. This is mainly because the dramatically increased input length not only significantly increases the context modeling burden of LLM rerankers, which degrades the ranking performance~\cite{fullrank}, but also leads to prohibitive inference latency~\cite{DBLP:conf/iclr/YueZB0JZ0WWB25}. Although prior work has explored rule-based strategies for processing long documents, such as truncation~\cite{DBLP:conf/sigir/DaiC19} and chunking~\cite{lrl, DBLP:journals/corr/abs-2312-10997, paptor,hipporag, graphlongdocument}, these approaches remain fundamentally limited. Truncation discards potentially important evidence appearing later in the document, while chunking breaks global semantic coherence across passages, both of which weaken the ranking model's ability to capture complete relevance signals.




To address these limitations, we propose \textbf{SumRank}, a novel summarization model oriented toward downstream long-document listwise reranking. Rather than just compressing text, this model is optimized to generate summaries that satisfy the strict requirements of the downstream ranking task. Specifically, we first deploy SumRank to compress each long document into a compact, query-aware summary that preserves essential relevance signals, and then perform listwise reranking over these summaries. SumRank is trained through a three-stage pipeline. We first initialize the model via cold-start supervised fine-tuning (SFT) using summaries distilled from a strong teacher model to learn basic document compression capabilities. We further align the summarizer with downstream ranking objectives via rank-driven reinforcement learning, where the reward is directly derived from the listwise ranking metric (NDCG@10) computed by an LLM-based reranker.

We evaluate the effectiveness of our SumRank on both in-domain (ID) TREC Deep Learning benchmarks (TREC DL 19–23) and a wide range of out-of-domain (OOD) datasets from BEIR. Experimental results demonstrate that our method achieves state-of-the-art (SOTA) ranking accuracy. By effectively compressing lengthy documents into summaries tailored for downstream ranking tasks, it consistently outperforms simple truncation methods, traditional summarizers, and zero-shot LLMs.

The main contributions of this work are: (1) We propose using a per-document summarization model before reranking to solve the challenges of long-document listwise reranking.  (2) We introduce a three-stage training framework, which includes a cold-start SFT strategy for basic summary generation, a specialized data construction strategy for the following RL, and a rank-driven alignment via an RL strategy to tailor the generated summaries specifically for the downstream ranking task. (3) We conduct extensive experiments to demonstrate the effectiveness and efficiency advantages of our SumRank.


\section{Related Work}

\subsection{Document Reranking with LLM}

With the advancement of LLMs, document reranking methodologies have evolved into three primary paradigms: pointwise, pairwise, and listwise. Pointwise methods~\cite{NogueiraJPL20,demorank} judge the relevance of a single query-document pair independently. Pairwise methods~\cite{QinJHZWYSLLMWB24} judge the relative relevance by comparing a pair of documents. Listwise approaches~\cite{rankgpt, lrl, rankvicuna, rankzephyr, coranking, fullrank, reasonrank, pe-rank} directly take a document list as the input, which captures the global document relevance, and output the reranked document IDs. While these methods are effective, they primarily focus on optimizing short passages reranking. Directly applying these methods to long-document tasks often leads to performance drops due to input length limitations and lack of generalizability.

\subsection{Long Document Modeling}
Due to the quadratic computational complexity of self-attention~\cite{DBLP:conf/nips/VaswaniSPUJGKP17}, standard Transformers are inefficient for long-document processing. To resolve this issue, a line of research focuses on optimizing the architecture by replacing the global fully connected attention with sparse attention mechanisms~\cite{cedr, colbert,longformer, bigbird, DBLP:conf/emnlp/JiangXL020, DBLP:conf/www/0002DYM22}. Other approaches tackle length constraints via fragmentation. Early methods relied on heuristic pooling (FirstP/MaxP/SumP)~\cite{DBLP:conf/sigir/DaiC19} or optimized training objectives based on cumulative gain~\cite{DBLP:conf/www/WuMLZZZM20} to aggregate passage scores. More recent works focus on efficiency and representation depth~\cite{DBLP:conf/sigir/HofstatterMZCH21,DBLP:journals/tois/LiYMHS24, li2025last,li2025enrich, li2026scaling}. However, these methods are primarily designed to accommodate long inputs, rather than optimizing for the downstream ranking objective. To bridge this gap, we propose SumRank, which directly compresses the long document into summaries tailored specifically for listwise ranking.



\section{Methodology}
\begin{figure*}[t]
    \centering
    \makebox[\textwidth]{%
    \includegraphics[trim={0.0cm 0.0cm 0.0cm 0.0cm}, clip, width=1\textwidth]{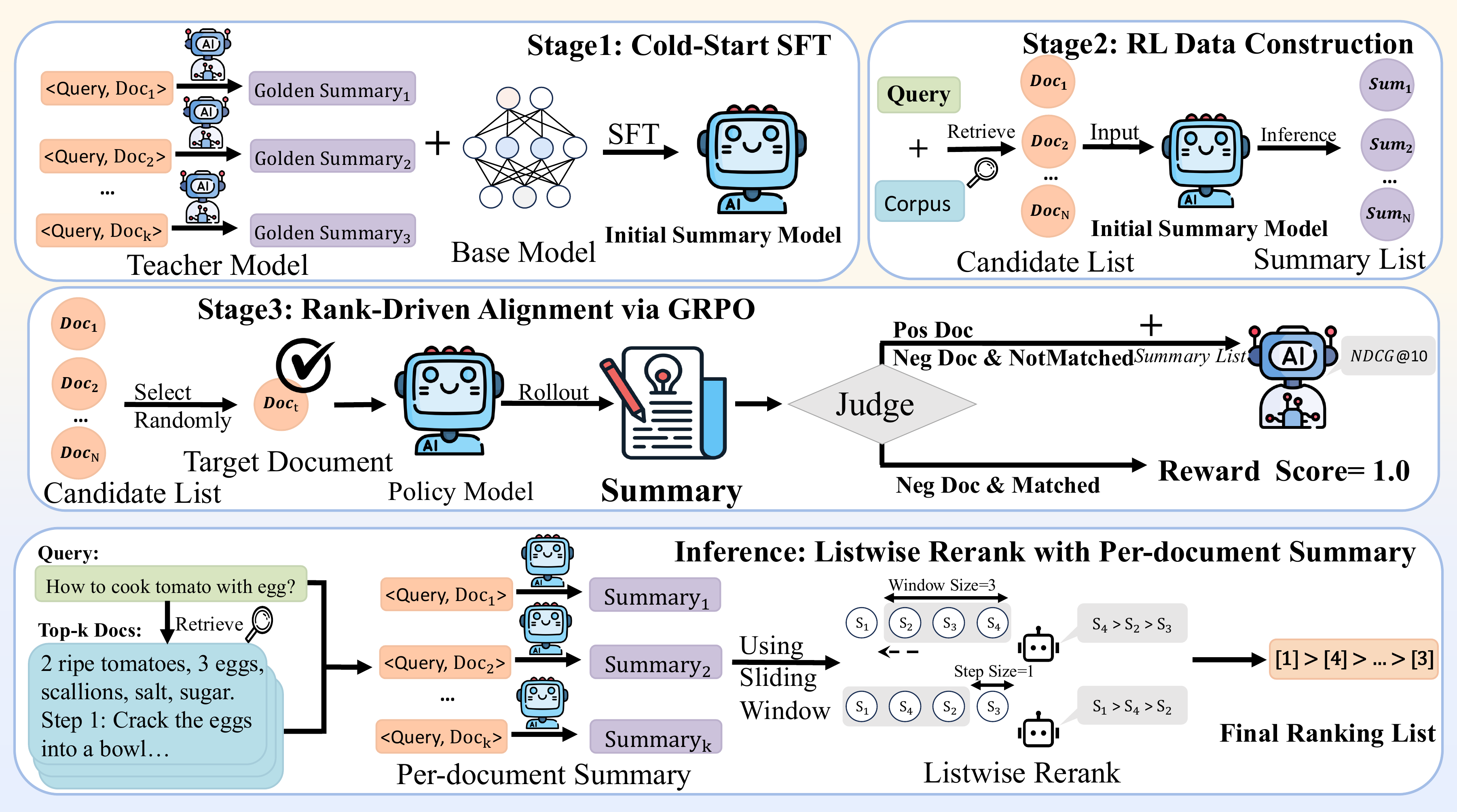}
    }
    \caption{The overall architecture of SumRank, which consists of three training stages and the Summary-then-Rank pipeline (Listwise Rerank with Per-document Summary).}
    \label{fig:workflow_trimmed}
\end{figure*}

In this section, we propose a comprehensive three-stage training framework to obtain a summarization model, \textbf{SumRank}, strictly aligned with downstream listwise reranking tasks. As illustrated in Figure~\ref{fig:workflow_trimmed}, this section is organized as follows: we first formulate the problem of reranking based on summarization (\S~\ref{sec:problrm formulation}). We then detail the three stages of our framework: Cold-Start SFT (\S~\ref{sec:Cold-Start SFT}), RL Data Construction(\S~\ref{sec:Data Construction}) and Rank-Driven Alignment via GRPO (\S~\ref{sec:Rank-Driven Alignment via GRPO}).

\subsection{Problem Formulation}
\label{sec:problrm formulation}

Given a query $q$ and a candidate document list $\mathcal{D} = \{d_1, ..., d_n\}$, the ultimate goal of document ranking is to determine a relevance permutation $\pi$ (\textit{i.e.}, [5] > [1] > …) that maximizes the relevance to the query $q$. 

To reduce computational cost and avoid the hallucination phenomenon of LLMs on long texts, we adopt the Summarize-then-Rank paradigm. Specifically, this paradigm introduces a per-document generative summarizer $\pi_\theta$ and a listwise ranker $R_\phi$. Each document $d_i$ is first compressed into a query-grounded summary: $s_i = \pi_\theta(d_i, q)$. Subsequently, the ranker $R_\phi$ evaluates the condensed summary set $\mathcal{S} = \{s_1, ..., s_n\}$ to output the final listwise ranking. Following prior work~\cite{rankgpt}, we employ a sliding window strategy (window size $w=20$, step size $sz=10$) to aggregate local rankings into the final global permutation $\pi$.

\subsection{Cold-Start SFT}
\label{sec:Cold-Start SFT}
To equip our model with preliminary summarization capabilities adapted to the downstream long-document listwise reranking task, we choose to distill the summarization generation ability of a strong teacher model into a lightweight student model parameterized by $\theta$. While powerful LLMs excel at generating high-quality summaries, deploying them directly incurs severe computational overhead~\cite{DBLP:journals/tacl/LiuLHPBPL24}. Therefore, we frame this initial stage as a knowledge distillation task~\cite{DBLP:journals/corr/HintonVD15, DBLP:conf/acl/HsiehLYNFRKLP23}.

Specifically, for each given query $q$ and long document $D$, we first use an instruction template $\mathcal{T}$ (detailed in~\cref{sec:prompt_templates}) to prompt the teacher LLM to generate a golden summary $y^*$. We then construct a standardized input prompt $X = \mathcal{T}(q, D)$ using the template $\mathcal{T}$. During the Supervised Fine-Tuning (SFT) phase, the student model is trained to replicate the teacher's output by minimizing the standard autoregressive cross-entropy loss:
\begin{equation}
\mathcal{L}_{\text{SFT}}(\theta) = - \frac{1}{T} \sum_{t=1}^{T} \log \pi_{\theta}(y_t^* | X, y_{<t}^*),
\end{equation}
where $y_{<t}^*$ denotes the sequence of prefix tokens generated prior to time step $t$, and $\pi_{\theta}$ represents the predicted probability distribution of the student model. By optimizing this token-level loss, our model efficiently internalizes the basic document compression capabilities, providing a solid initialization for the subsequent reinforcement learning stage.

\subsection{RL Data Construction}
\label{sec:Data Construction}

To construct the environment for our reinforcement learning stage, we construct a standard training dataset $\mathcal{D}$ containing queries with corresponding positive and negative documents. To facilitate listwise evaluation, we construct a candidate list $\mathcal{L}_D$ of size $N$.

To guarantee a valid listwise learning signal, the evaluation list must simultaneously contain both positive and negative documents. We first employ an initial retriever (\eg, BM25) to obtain a top-$N$ candidate list $\mathcal{R}_q$. We then evaluate the composition of $\mathcal{R}_q$: if it already includes both positive and negative instances, we utilize it directly. Otherwise, if $\mathcal{R}_q$ lacks either positive document $\mathcal{D^+}$ or negative document, we explicitly inject $k$ missing documents $\mathcal{D}^*_k$ (sampled from the ground-truth positive or negative set, respectively) by replacing the lowest-ranked candidates in $\mathcal{R}_q$:
\begin{equation}
  \mathcal{L}_D = 
 \mathrm{Shuf}(\mathcal{R}_q[1 : N - k] \cup \mathcal{D}^*_k).
\end{equation}

Finally, we utilize the cold-started student model $\pi_{\text{SFT}}$ to decode an initial summary for every document in $\mathcal{L}_D$, yielding a static background summary list $\mathcal{L}_S$:
\begin{equation}
\label{eq:static_list}
  \mathcal{L}_S = \big\{ s_i \mid s_i \sim \pi_{\text{SFT}}(\cdot | X_i), \forall D_i \in \mathcal{L}_D \big\}.
\end{equation}

\subsection{Rank-Driven Alignment via GRPO}
\label{sec:Rank-Driven Alignment via GRPO}

With the static summary list $\mathcal{L}_S$ established, we optimize the cold-start summarization model $\pi_{\text{SFT}}$ using Group Relative Policy Optimization (GRPO)~\cite{DBLP:journals/corr/abs-2402-03300}.

\subsubsection{Target Selection and Rollout}
During training for a given query $q$ and its associated candidate document list $\mathcal{L}_D$, we randomly select a single target document $D_t$ from the N-document list $\mathcal{L}_D$. Due to our data construction mechanism, $D_t$ could be either a positive or a negative document.

Once $D_t$ is selected, we formulate the input prompt $X_t = \mathcal{T}(q, D_t)$. To construct the comparison group required by GRPO, we prompt the active policy to auto-regressively sample a group of $G$ candidate summary rollouts, denoted as $Y = \{y_1, y_2, \dots, y_G\}$. Each rollout $y_i \in Y$ represents a variant action taken by the model to condense $D_t$.

\subsubsection{Listwise Evaluation and Reward}
The reward mechanism evaluates how every single rollout $y_i$ affects the global ranking of its corresponding document. To evaluate a specific rollout $y \in Y$ with minimal computational overhead, we construct a mixed evaluation list $\hat{\mathcal{L}}_S$. We replace the $t$-th static summary $s_t$ (representing $D_t$) in the pre-generated static Summary List $\mathcal{L}_S$ with the newly generated rollout $y$, while keeping the other $N-1$ summaries fixed. Formally, the assembled list is $\hat{\mathcal{L}}_S = \{s_1, \dots, s_{t-1}, y, s_{t+1}, \dots, s_{N}\}$. 

This design choice is fundamentally driven by two critical motivations: reducing reward variance and ensuring computational efficiency.
\textbf{First}, listwise ranking evaluates relative relevance. By fixing the $N-1$ non-target summaries using the baseline $\pi_{\text{SFT}}$ model, we establish a stable reference frame. Forcing the model to compete against the baseline $\pi_{\text{SFT}}$ model's capabilities ensures that any rank improvement reflects a genuine gain in summarization quality, rather than environmental noise.
\textbf{Second}, generating an entire $N$-document list for the $G$ sampled rollouts required by GRPO would lead to cost overhead. The static background list reduces this generative cost from $\mathcal{O}(N \times G)$ to $\mathcal{O}(G)$ per step, making listwise RL practically feasible for long-document tasks.

An LLM reranker then performs listwise reranking on this updated N-document list. After the downstream listwise ranker scores and sorts the candidates in $\hat{\mathcal{L}}_S$, the reward for the generated summary $y$ is computed.
We define the reward signal $R(y)$ using the standard Normalized Discounted Cumulative Gain (NDCG@$10$) for the listwise reranking results.  By directly optimizing this ranking metric, the policy model learns to generate summaries that inherently maximize the comparative prominence of relevant documents.

To prevent the model from manufacturing false alignments to irrelevant candidates, we introduce a lexical matching indicator $M(y) \in \{0, 1\}$. Specifically, $M(y)=1$ if the output $y$ contains the safeguard phrase ``No Relevant Information Found'' (as explicitly instructed in our prompt template $\mathcal{T}$) or its closely related semantic variants. 

The comprehensive reward dynamically conditions on the ground-truth label of the target document $D_t$:
\begin{equation}
\resizebox{0.88\hsize}{!}{$
  r(y) = 
  \begin{cases} 
    R(y), & \text{if } D_t = D^+, \\ 
    1.0, & \text{if } D_t \neq D^+ \land M(y) = 1, \\ 
    R(y) - \lambda, & \text{if } D_t \neq D^+ \land M(y) = 0,
  \end{cases}
$}
\end{equation}
where $\lambda$ is a penalty factor applied when the model fails to explicitly reject a negative document but still undergoes ranking evaluation.

\subsubsection{Optimization Objective}

Finally, the reward $r_i$ for each rollout $y_i$ is Z-score normalized within the group to yield the relative advantage $\hat{A}_i = (r_i - \mu_Y) / \sigma_Y$. The active policy $\pi_{\theta}$ is then optimized by maximizing the streamlined GRPO objective:
\begin{equation}
\begin{aligned}
\mathcal{J}_{\text{GRPO}}(\theta) &= \frac{1}{|G|} \sum_{i=1}^{|G|} \frac{1}{|y_i|} \sum_{t=1}^{|y_i|} \min \Big( \rho_{i,t}(\theta) \hat{A}_{i,t},\\
   &\quad \text{clip}\big(\rho_{i,t}(\theta), 1 \pm \epsilon\big) \hat{A}_{i,t} \Big  ) - \beta D_{\text{KL}},
\end{aligned}
\end{equation}
where $|y_i|$ is the sequence length of $y_i$, $\epsilon$ is the clipping threshold, and $\beta$ is the scaling coefficient for the KL divergence penalty. 

$\rho_{i,t}(\theta)$ denotes the token-level probability ratio. Crucially, $D_{\text{KL}}$ represents the token-level Kullback-Leibler divergence between the active policy $\pi_\theta$ and the cold-start reference model $\pi_{\text{SFT}}$, which effectively anchors the optimization space and prevents mode collapse. These two components are formally defined as follows:
\begin{equation}
\begin{aligned}
  \rho_{i,t}(\theta) &= \frac{\pi_\theta(y_{i,t} \mid X, y_{i,<t})}{\pi_{\text{old}}(y_{i,t} \mid X, y_{i,<t})}, \\
  D_{\text{KL}} &= D_{\text{KL}} \big( \pi_\theta \parallel \pi_{\text{SFT}} \big).
\end{aligned}
\end{equation}

\begin{table*}[t]
\centering
\resizebox{\textwidth}{!}{
\begin{tabular}{l ccccc c c ccccc c}
\toprule
\multirow{3}{*}{\textbf{Methods}} & \multicolumn{6}{c}{\textbf{In-Domain (ID)}} & & \multicolumn{6}{c}{\textbf{Out-of-Domain (OOD)}} \\
\cmidrule(lr){2-7} \cmidrule(lr){9-14}
& \textbf{DL 19} & \textbf{DL 20} & \textbf{DL 21} & \textbf{DL 22} & \textbf{DL 23} & \textbf{Avg.} & & \textbf{covid} & \textbf{news} & \textbf{nfcorpus} & \textbf{robust04} & \textbf{scifact} & \textbf{Avg.} \\
\midrule
BM25 & 51.76 & 52.86 & 51.16 & 29.93 & 29.46 & 43.03 & & 59.47 & 39.52 & 33.75 & 40.70 & 67.89 & 48.27 \\
\midrule
\multicolumn{14}{l}{\textit{Rule-based Methods}} \\
FirstP-128 & 64.41 & 61.60 & 66.84 & 40.82 & 42.39 & 55.23 & & 74.96 & 41.22 & 35.85 & 50.37 & 64.63 & 53.41 \\
FirstP-256 & 65.22 & 61.07 & 67.56 & 40.66 & 41.58 & 55.15 & & 75.36 & 42.35 & 34.60 & 49.44 & 66.40 & 53.63 \\
FirstP-512 & 65.94 & 61.25 & 67.20 & 39.82 & 41.14 & 55.07 & & 75.78 & 42.41 & 34.12 & 50.80 & 65.32 & 53.69 \\
maxP & 65.59 & 62.71 & 66.71 & 39.81 & 38.96 & 54.76 & & 77.81 & 42.76 & 35.21 & 48.36 & 66.14 & 54.06 \\
\midrule
\multicolumn{14}{l}{\textit{Seq2Seq Summarization Models}} \\
BART-summarizer & 62.05 & 61.48 & 67.53 & 40.07 & 41.66 & 54.96 & & 69.81 & 40.14 & 37.07 & 43.50 & 66.07 & 51.32 \\
LED-summarizer & 59.80 & 57.82 & 67.22 & 40.18 & 41.36 & 53.28 & & 67.64 & 37.25 & 35.75 & 48.55 & 72.15 & 52.27 \\
PEGASUS & 62.02 & 60.83 & 67.53 & 40.46 & 41.84 & 54.54 & & 69.35 & 43.68 & 36.90 & 44.57 & 69.85 & 52.87 \\
\midrule
\multicolumn{14}{l}{\textit{LLM-based Summarization Models}} \\
Qwen2.5-3B-Instruct & 58.97 & 54.45 & 63.67 & 37.17 & 35.52 & 49.96 & & 75.75 & 41.41 & 35.91 & 47.84 & 70.68 & 54.32 \\
Qwen2.5-7B-Instruct & 59.72 & 62.03 & 66.15 & 35.40 & 40.24 & 52.71 & & 78.30 & 41.43 & 35.55 & 48.68 & 70.88 & 54.97 \\
\midrule
\multicolumn{14}{l}{\textit{Our Methods}} \\
SumRank (3B) & \underline{67.06} & \textbf{62.99} & \underline{68.09} & \underline{42.22} & \underline{43.76} & \underline{56.82} & & \underline{81.92} & \underline{47.24} & \underline{37.26} & \underline{51.38} & \underline{72.25} & \underline{58.01} \\
SumRank (7B) & \textbf{67.30} & \underline{62.95} & \textbf{69.30} & \textbf{42.74} & \textbf{44.31} & \textbf{57.32} & & \textbf{81.69} & \textbf{47.44} & \textbf{38.27} & \textbf{54.71} & \textbf{72.91} & \textbf{59.00} \\
\bottomrule
\end{tabular}
}
{\fontsize{9.5pt}{11.5pt}\selectfont
\caption{The NDCG@10 results on TREC DL 19-23 (In-Domain) and BEIR (Out-of-Domain) benchmarks. All baselines rerank the BM25-retrieved top-100 documents. The best results are highlighted in \textbf{bold}, and the second-best results are \underline{underlined}.}
\label{tab:main_results_updated}
}
\end{table*}

\section{Experimental Setup}

\subsection{Datasets and Metrics}

We comprehensively evaluate our framework across diverse, real-world retrieval scenarios using the standard \textbf{TREC Deep Learning} (DL) tracks across five datasets: TREC DL19~\cite{dl19}, TREC DL20~\cite{dl20}, TREC DL21~\cite{dl21}, TREC DL22~\cite{dl22}, and TREC DL23~\cite{dl23}. 

Furthermore, to demonstrate the temporal robustness and consistent generalization of SumRank, we extend our evaluation to five Out-of-Domain (OOD) datasets from the BEIR benchmark~\cite{beir}, spanning biomedical search (\textit{TREC-COVID}, \textit{SciFact}), web search (\textit{Robust04}), and dedicated text domains (\textit{TREC-News}, \textit{NFCorpus}).

To balance computational cost and ranking quality, for the downstream listwise reranking phase, we employ the highly capable \textbf{Qwen2.5-32B-Instruct}~\cite{DBLP:journals/corr/abs-2412-15115} as our document reranker. To maximize overall ranking performance, we adopt the sliding window inference strategy~\cite{rankgpt} setting the window size to 20 and the step size to 10. We employ NDCG@10 as the evaluation metric across all benchmarks.

\subsection{Baselines}

We compare with three types of baselines.

\paragraph{Rule-based Methods}
We utilize BM25 to establish the initial ranking, alongside the widely adopted FirstP~\cite{DBLP:conf/sigir/DaiC19} strategy, which truncates documents to their first $k$ tokens ($k \in \{128, 256, 512\}$) to leverage document ``lead bias''. We also evaluate the MaxP~\cite{DBLP:conf/sigir/DaiC19} strategy that determines the document relevance based on its highest-scoring chunk.

\paragraph{Seq2Seq Summarization Models}
We benchmark against sequence-to-sequence models explicitly pre-trained and fine-tuned for abstractive summarization tasks. This group includes the foundational BART-summarizer~\cite{DBLP:conf/acl/LewisLGGMLSZ20}, its long-document sparse-attention extension LED-summarizer~\cite{longformer}, and PEGASUS~\cite{DBLP:conf/icml/ZhangZSL20}, which is specifically optimized for summarization generation. 

\paragraph{LLM-based Summarization Models} 
To assess the raw summarization capabilities of modern foundational models without our proposed listwise alignment, we employ the instruction-tuned Qwen2.5-3B-Instruct and Qwen2.5-7B-Instruct~\cite{DBLP:journals/corr/abs-2412-15115}. We prompt these models to directly generate document summaries.

\subsection{Implementation Details}

For the Cold-Start SFT stage, we employ LLaMA-Factory~\cite{DBLP:conf/acl/ZhengZZYL24} to perform full-parameter fine-tuning on Qwen2.5-3B-Instruct and Qwen2.5-7B-Instruct. Both models are trained on a 40k-instance distillation dataset, derived from the MS MARCO dataset~\cite{DBLP:conf/nips/NguyenRSGTMD16} and generated by the teacher model, Qwen2.5-72B-Instruct. This stage is trained for 5 epochs with a learning rate of $5 \times 10^{-6}$ using 4 NVIDIA A800 (80GB) GPUs.

For the Data Construction stage, we sample 2,500 queries from MS MARCO~\cite{DBLP:conf/nips/NguyenRSGTMD16} to construct 2,500 corresponding candidate document groups. We set N to 10 and k to 1. During the subsequent GRPO training, the policy rollout for each query group is executed with one positive document and one negative document.

For the Rank-Driven Alignment via GRPO stage, we employ the verl framework~\cite{sheng2024hybridflow} for reinforcement learning. We initialize the policy with the SFT checkpoint. We train the 3B model on two NVIDIA A800 (80GB) GPUs and the 7B model on four NVIDIA A800 (80GB) GPUs, both using a reduced learning rate of $1 \times 10^{-6}$. We sample $G=8$ rollouts per prompt, train for 5 epochs, and set the KL divergence penalty $\beta=0.001$. We employ Qwen2.5-72B-Instruct as the listwise LLM reranker to compute the NDCG@10 reward, with the penalty $\lambda$ set to 0.25.

\section{Experimental Results}

\subsection{Main Results}

The main evaluation results on both In-Domain (TREC DL 19-23) and Out-of-Domain (BEIR) benchmarks are summarized in Table \ref{tab:main_results_updated}. SumRank (7B) achieves state-of-the-art performance across all datasets, with SumRank (3B) closely following and consistently surpassing all baselines.

\paragraph{Comparison with Rule-based Methods} 
Naive truncation strategies (\eg, FirstP-256) establish surprisingly strong baselines on the In-Domain (ID) benchmarks. This is primarily because TREC DL datasets take advantage of the ``lead bias'', which means the tendency for web pages to put key information at the very beginning. However, the effectiveness of truncation degrades significantly on datasets with different structural distributions (\eg, biomedical documents), where crucial evidence is often deeply buried. SumRank solves this problem by compressing the entire document. Instead of just looking at the beginning, it successfully captures the valuable evidence from the middle and end of the text that truncation simply throws away.

\paragraph{Comparison with Seq2Seq Summarization Models} 
The results of classic summarizers (BART-summarizer, LED-summarizer, PEGASUS) clearly show that their original goals do not match the ranking task. Since these models are trained to write smooth, general overviews for human readers, they accidentally delete the specific keywords and exact details that a ranking model desperately needs to calculate relevance.

\paragraph{Comparison with LLM-based Summarization Models}
The comparison between SumRank (3B) and the much larger Qwen2.5-7B-Instruct highlights that zero-shot summaries generated by general LLMs simply cannot meet the strict requirements of downstream ranking. However, once our 3B model undergoes the cold-start SFT and rank-driven alignment via the GRPO training pipeline, its ranking performance improves dramatically, easily beating the 7B baseline. This clearly shows that specifically training a smaller model to preserve relevance is far more powerful than relying on the zero-shot capabilities of a larger model.

\subsection{Ablation Study}

To investigate the contribution of different training strategies, we conducted an ablation study by removing the cold-start SFT stage and rank-driven alignment via the GRPO stage independently. As shown in Table~\ref{tab:comprehensive_ablation}, the full model consistently achieves the best performance across the TREC DL dataset, confirming that both components are indispensable.

When we remove the rank-driven alignment via the GRPO stage (\ie, ``\textit{w/o} rank-driven GRPO ''), the average score suffers the biggest drop. This proves that relying on SFT alone is not enough. While SFT is great at teaching the model the basic format and length of a summary, it cannot fully align the model with the final ranking goal. Without the targeted rewards provided by rank-driven alignment via GRPO, the model easily slips back into generating generic text and forgets to highlight the specific keywords needed for reranking.

Removing the initial SFT phase (\ie, ``\textit{w/o} cold-start SFT'') also hurts the final results. This happens because reinforcement learning needs a solid starting point. Without SFT to first show the model exactly what a proper summary should look like, the GRPO training process has a much harder time figuring out the right direction from scratch, leading to less stable and suboptimal learning.

\begin{table}[htbp]
\centering
\small 
\begin{tabular}{lcc}
\toprule
\textbf{Model Variant} & \textbf{TREC Avg.} & \textbf{$\Delta$} \\
\midrule
SumRank (3B) & \textbf{56.82} & - \\
\midrule
\multicolumn{3}{l}{\textit{Training Approach}} \\
\textbullet~\textit{w/o} Cold-Start SFT & 55.88 & -0.94 \\
\textbullet~\textit{w/o} Rank-Driven GRPO & 54.92 & -1.90 \\
\midrule
\multicolumn{3}{l}{\textit{Data Construction}} \\
\textbullet~Truncation FirstP128 & 54.88 & -1.94 \\
\textbullet~Truncation FirstP256 & 54.92 & -1.90 \\ 
\textbullet~Update per epoch & 56.32 & -0.50 \\
\bottomrule
\end{tabular}
{\fontsize{9.5pt}{11.5pt}\selectfont
\caption{Ablation study on Training Approach and Data Construction. ``TREC Avg.'' reports the average NDCG@10 scores evaluated across all five TREC DL benchmarks (DL19--23). $\Delta$ indicates the absolute performance degradation on specific strategies.}
\label{tab:comprehensive_ablation}
}
\end{table}

Ultimately, the full pipeline achieves the highest average score. This confirms that the cold-start SFT stage and the rank-driven alignment via the GRPO stage perfectly complement each other. SFT builds the necessary foundation, and GRPO acts as the crucial final step to lock in those vital matching signals for the downstream reranking.

As presented in Table~\ref{tab:comprehensive_ablation}, we investigate the impact of different data construction strategies for representing non-target documents.

Replacing our generated summaries with simple truncated raw texts (\ie, ``Truncation FirstP128/256'') leads to a performance drop. Notably, whether we truncate at 128 or 256 tokens, the degradation remains equally severe. This clearly demonstrates that naive truncation is fundamentally ineffective. Raw document prefixes are dominated by noise and fail to provide the policy model with relevance signals.

\begin{table*}[t]
\centering
\resizebox{\textwidth}{!}{
\begin{tabular}{l cc cc cc cc cc cc}
\toprule
\multirow{2}*{\textbf{Method}} & \multicolumn{2}{c}{\textbf{TREC DL 19}} & \multicolumn{2}{c}{\textbf{TREC DL 20}} & \multicolumn{2}{c}{\textbf{TREC DL 21}} & \multicolumn{2}{c}{\textbf{TREC DL 22}} & \multicolumn{2}{c}{\textbf{TREC DL 23}} & \multicolumn{2}{c}{\textbf{Avg.}} \\
\cmidrule(lr){2-3} \cmidrule(lr){4-5} \cmidrule(lr){6-7} \cmidrule(lr){8-9} \cmidrule(lr){10-11} \cmidrule(lr){12-13}
& Rate & NDCG & Rate & NDCG & Rate & NDCG & Rate & NDCG & Rate & NDCG & Rate & NDCG \\
\midrule
\multicolumn{13}{l}{\textit{SumRank (3B)}} \\
\quad w/ rejection  & 86.79\% & 67.06 & 90.64\% & 62.99 & 84.33\% & 68.09 & 90.55\% & 42.22 & 92.35\% & 43.76 & 88.93\% & 56.82 \\
\quad w/o rejection & 87.70\% & 65.23 & 91.07\% & 61.12 & 84.72\% & 68.09 & 91.42\% & 42.50 & 92.91\% & 42.71 & 89.56\% & 55.93 \\
\midrule
\multicolumn{13}{l}{\textit{SumRank (7B)}} \\
\quad w/ rejection  & 85.65\% & 67.30 & 90.76\% & 62.95 & 80.35\% & 69.30 & 88.91\% & 42.74 & 91.94\% & 44.31 & 87.52\% & 57.32 \\
\quad w/o rejection & 82.30\% & 67.25 & 88.00\% & 64.82 & 76.26\% & 69.92 & 87.22\% & 43.37 & 90.28\% & 44.90 & 84.81\% & 58.05 \\
\bottomrule
\end{tabular}
}
{\fontsize{9.5pt}{11.5pt}\selectfont
\caption{Analysis of the impact of the explicit rejection prompt, ``No relevant information found''. ``w/ rejection'' denotes the standard prompt advising the model to explicitly output "No relevant information found" if the document is irrelevant, while ``w/o rejection'' completely removes this hard-coded instruction. The \textbf{Rate} indicates the proportion of summaries as ``No relevant information found''. The \textbf{NDCG} indicates the NDCG@10 results of downstream ranking performance.}
\label{tab:prompt_ablation_final}
}
\end{table*}

Furthermore, dynamically updating the context summaries at every training epoch (\ie, ``Update per epoch''), instead of using our default static summaries, unexpectedly decreases the performance. This confirms that keeping the context summaries static provides a stable and consistent reference frame for the model during RL training, whereas a constantly shifting background introduces instability. More importantly, regenerating summaries every epoch incurs computational overhead. Therefore, our simple static strategy proves to be both more efficient and more effective.

\subsection{Efficiency Analysis}

\begin{figure}[htbp]
\centering
\includegraphics[width=\columnwidth]{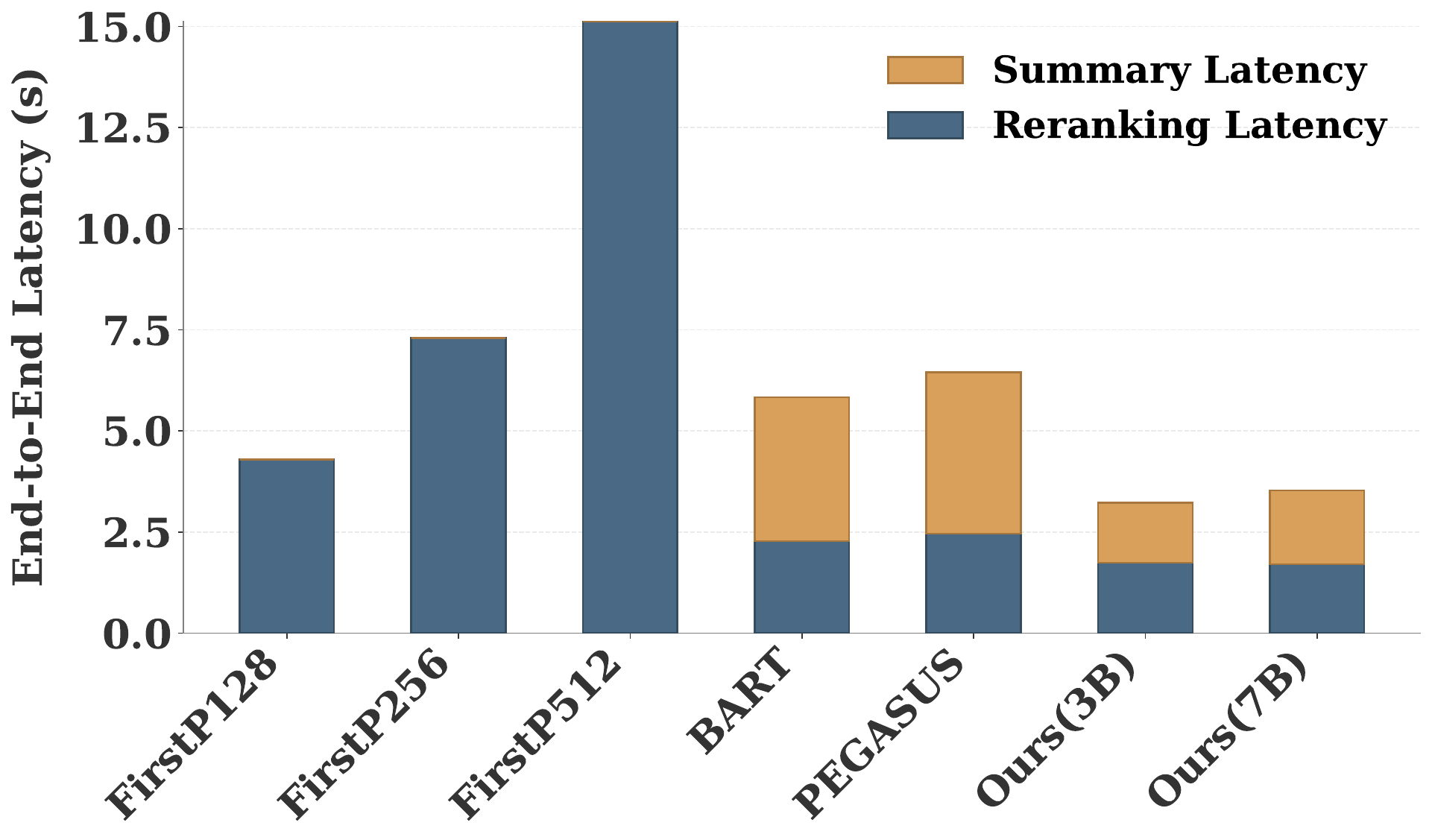} 
\caption{Decomposition of end-to-end inference latency on TREC DL benchmarks. The total processing time is explicitly broken down into \textit{Summary Generation Latency} (measured in seconds/query) and \textit{Listwise Reranking Latency} (measured in seconds/query).}
\label{fig:latency_decomposition}
\end{figure}

To comprehensively evaluate the practical value of our framework, we investigate how SumRank and the baseline models impact the overall end-to-end inference efficiency. Figure~\ref{fig:latency_decomposition} decomposes the total processing time into Summary Generation Latency and Listwise Reranking Latency.

Truncation-based methods (\eg, FirstP512) completely avoid generation overhead but suffer from listwise reranking bottleneck. This massive delay is caused by the quadratic complexity of the ranker's self-attention mechanism when processing long raw texts. Traditional Seq2Seq models (\eg, LED, BART) alleviate the reranking burden through text compression, yet their massive generation overheads destroy potential gains in end-to-end efficiency.

SumRank directly resolves this trade-off. It reduces lengthy texts into dense, rank-aligned summaries, which significantly accelerates downstream listwise reranking. Because this compression introduces minimal generation delay, SumRank achieves an exceptional end-to-end latency that is faster than both traditional truncation and summarization baselines. By eliminating the efficiency bottleneck, SumRank balances between computational speed and ranking performance.

\subsection{Analysis of Explicit Rejection}

To investigate whether SumRank's capability to reject negative examples stems merely from prompt injection or is acquired through the rank-driven alignment stage, we conduct an ablation study. Specifically, we evaluate the model's rejection rate and downstream ranking performance after completely removing the hard-coded rejection instruction (\eg, ``output 'No relevant information found' if irrelevant'') from the inference prompt. The results are presented in Table \ref{tab:prompt_ablation_final}.

Even without explicit instructions, both the 3B and 7B models autonomously generate the vast majority of irrelevant documents. This confirms that their capability to reject negative examples is not from prompt engineering, but learned through the rank-driven alignment stage.

Furthermore, removing the prompt reveals a divergence in downstream ranking based on model capacity. For the smaller SumRank (3B), the absence of explicit instructions leads to a slight performance degradation. This suggests that smaller models still require a structural prompt to maintain stability. In contrast, the 7B model exhibits improved performance without the prompt, indicating that the larger model has sufficiently internalized relevance alignment through the reinforcement learning stage.

\section{Conclusion}

We present SumRank, a summarization model aligned with downstream document listwise reranking. We designed a tailored three-stage training paradigm to synthesize high-density relevance features and utilize an LLM reranker for rank-aware preference alignment. Through this well-structured training process, SumRank effectively provides the downstream ranker with the exact matching signals extracted from the original long documents, achieving a significant improvement in ranking performance while reducing end-to-end listwise reranking latency.


\section*{Limitations}

Although SumRank significantly improves the efficiency of long-document reranking, it has certain limitations. Firstly, due to our limited resources, we cannot experiment with larger open-source LLMs, such as 14B or even 32B models. It is interesting to investigate the alignment of larger-scale models. Secondly, our approach relies on a two-stage "Summary-then-Rank" pipeline. This means we still need to deploy and maintain two separate models: one for summarization and another for ranking. In the future, we plan to explore a unified, end-to-end architecture. We aim to train a single model that can simultaneously compress the document and output the final relevance score, which will further simplify deployment and reduce inference latency.

\section*{Acknowledge}
This work is supported by the National Natural Science Foundation of China No.62272467. 


\appendix

\section{Use of AI Assistants}
\label{sec:appendix}
We use Gemini\footnote{https://gemini.google.com/} to improve the presentations of this paper.

\section{Prompt Templates}
\label{sec:prompt_templates}

To ensure the reproducibility of our experiments and guide the large language models (LLMs) effectively, we design two distinct prompt templates for the distinct stages of our pipeline: \textbf{Per-document Summarization} and \textbf{Listwise Reranking}. The comprehensive formulations of these templates are illustrated below.

\vspace{0.3cm}

\begin{tcolorbox}[
    colback=black!5!white,      
    colframe=black!75!white,    
    coltitle=white,             
    fonttitle=\bfseries,        
    title=Prompt for Per-document Summarization., 
    arc=2mm,                    
    boxrule=1pt,                
    left=3mm, right=3mm, top=2mm, bottom=2mm
]
Task: Generate the summary from the Document that answers the Query.\vspace{0.5\baselineskip}

Strict Rules:
\begin{enumerate}[topsep=0pt, itemsep=0pt, parsep=0pt, partopsep=0pt, leftmargin=*]
    \item If the Document contains the answer, extract it word-for-word or summarize it faithfully.
    \item If the Document is \textbf{NOT} relevant to the Query, you \textbf{MUST} output exactly: 'No relevant information found.'
    \item Do \textbf{NOT} provide explanations, apologies, or conversational fillers like 'The document says...'. Just the answer or the refusal phrase.
\end{enumerate}
\vspace{0.5\baselineskip}
Query: \{query\} \\Document: \{document\}
\end{tcolorbox}
\begin{tcolorbox}[
    colback=black!5!white,      
    colframe=black!75!white,    
    coltitle=white,             
    fonttitle=\bfseries,        
    title=Prompt for Listwise Rerank., 
    arc=2mm,                    
    boxrule=1pt,                
    left=3mm, right=3mm, top=2mm, bottom=2mm 
]

I will provide you with \{N\} passages, each indicated by a numerical identifier []. Rank the passages based on their relevance to the search query: \{query\}.\vspace{0.5\baselineskip}

[1] \{document1\}

[2] \{document2\}

……

[N] \{documentN\}\vspace{0.5\baselineskip}

Search Query: \{query\}.\vspace{0.5\baselineskip}

Rank the \{N\} passages above based on their relevance to the search query. All the passages should be included and listed using identifiers, in descending order of relevance. The output format should be [] > [], e.g., [2] > [1], Only respond with the ranking results, do not say any word or explain.
\end{tcolorbox}

\section{Summary Length Statistics}

\begin{table*}[t]
\centering
\small
\begin{tabular}{l ccccc c ccccc}
\toprule
\textbf{Metric} & \multicolumn{5}{c}{\textbf{TREC DL (In-Domain)}} & & \multicolumn{5}{c}{\textbf{BEIR (Out-of-Domain)}} \\
\cmidrule(lr){2-6} \cmidrule(lr){8-12}
& \textbf{DL19} & \textbf{DL20} & \textbf{DL21} & \textbf{DL22} & \textbf{DL23} & & \textbf{covid} & \textbf{news} & \textbf{nfcorpus} & \textbf{robust04} & \textbf{scifact} \\
\midrule
Original & 4193 & 4384 & 6542 & 5513 & 6096 & & 329 & 1472 & 353 & 2043 & 352 \\
\midrule
\multicolumn{12}{l}{\textit{SumRank (3B)}} \\
\quad w/o n & 140.04 & 88.10  & 103.44 & 153.35 & 160.15 & & 92.74 & 96.68  & 115.88 & 124.01 & 73.66 \\
\quad w n   & 22.87  & 12.80  & 20.49  & 19.09  & 16.87  & & 35.90 & 12.18  & 13.11  & 10.67  & 5.71  \\
\midrule
\multicolumn{12}{l}{\textit{SumRank (7B)}} \\
\quad w/o n & 135.56 & 100.68 & 106.81 & 180.17 & 170.80 & & 117.89 & 177.45 & 145.19 & 187.52 & 113.03 \\
\quad w n   & 23.73  & 13.88  & 25.05  & 24.44  & 18.37  & & 50.11  & 16.35  & 22.40  & 17.32  & 5.98  \\
\bottomrule
\end{tabular}
\caption{Average token counts of original documents and generated summaries across TREC DL (In-Domain) and BEIR (Out-of-Domain) benchmarks. The notation (w/o n) indicates the average token count of the generated summaries excluding the instances outputted ``No relevant information found,'' while (w n) includes all instances.}
\label{tab:combined_summary_lengths}
\end{table*}

Table~\ref{tab:combined_summary_lengths} presents a comprehensive statistical analysis of the average token lengths for both the original candidate documents and the summaries generated by \mbox{SumRank} across the TREC DL and BEIR benchmarks. Crucially, these statistics reveal several noteworthy insights regarding the behavior and efficiency of our proposed framework.

The summaries generated by \mbox{SumRank} under the \textit{w/o n} setting (excluding ``No relevant information found.'' instances) represent an explicit summary compression from the long document. The average token count under the \textit{w n} setting drops drastically. This reduction show that the strict refusal mechanism (\textit{``No relevant information found.''}) accurately identifies and filters out a vast majority of irrelevant or noisy candidate documents in the initial phase.

Comparing the 3B and 7B variants, \mbox{SumRank} (7B) generally produces slightly longer and richer summaries under the \textit{w/o n} setting. This indicates that the larger model possesses a stronger capacity to capture comprehensive evidence from lengthy contexts, without compromising the overall efficiency.

Ultimately, this remarkable compression capability directly addresses the core bottleneck of downstream listwise reranking. The original documents typically contain thousands of tokens, which poses a heavy computational burden for LLM-based listwise rankers. By inputting highly condensed, informative summaries into the listwise reranker instead of raw documents, \mbox{SumRank} drastically minimizes the overall sequence length. This reduction effectively alleviates computational bottlenecks in terms of inference latency.

\clearpage             
\onecolumn

\section{Case Study}

In this section, we first compare the summaries generated by SumRank (7B) and zero-shot Qwen2.5-7B-Instruct (Table \ref{tab:qualitative_analysis}). We randomly sample qualitative analysis over 5 query-document pairs from the TREC DL datasets, covering both relevant and irrelevant documents. 

The examples show that SumRank does not simply produce keyword summaries. Instead, it tends to preserve query-specific relevance evidence, retain important details, and reject irrelevant documents. In contrast, the zero-shot Qwen2.5-7B summarizer sometimes produces fluent but partially biased summaries or extracts topically related but non-answering content.

Additionally, we provide concrete examples of listwise reranking using generated summaries. We provide a detailed example of the summary output generated by SumRank (Table \ref{tab:sumrank_output_example_single}), followed by a detailed case study illustrating the listwise reranking process across 20 summaries (Table \ref{tab:listwise_output_example}).


\begin{longtable}{>{\bfseries}p{0.15\textwidth} p{0.8\textwidth}}
\caption{Comparison of generated summaries between SumRank (7B) and Qwen2.5-7B on sampled TREC DL query-document pairs.}
\label{tab:qualitative_analysis} \\
\toprule
\endfirsthead
\multicolumn{2}{c}%
{{\bfseries \tablename\ \thetable{} -- continued from previous page}} \\
\toprule
\endhead
\bottomrule
\endfoot
\bottomrule
\endlastfoot

\rowcolor{gray!15}
Query & \textbf{is levothyroxine likely to cause weight loss or weight gain} \\
\midrule
SumRank(7B) & Regular administration of synthroid (levothyroxine) is thought to reverse the symptoms of hypothyroidism and improve functionality of the user. \textbf{After several months of properly-dosed synthroid, users often report weight loss.} Weight loss is often among the most discussed synthroid side effects due to the fact that nearly everyone wants to look good and maintain a healthy physique. Additionally, many individuals had gained a significant amount of weight prior to using synthroid (as a result of hypothyroidism), a condition which slows metabolism. When metabolic processes are corrected as a result of synthroid, users end up losing weight and the weight loss comes as a pleasant surprise. \textbf{However, the scientific literature suggests that weight loss is unlikely.} A study by Lee et al. (2014) indicated that no significant changes in weight occur after initiation of levothyroxine (LT4) treatment. Another study by Ozdemir et al. (2010) suggested that levothyroxine may cause weight gain in older adults (ages 45+). \\
\addlinespace
Qwen2.5-7B & Synthroid (Levothyroxine) and Weight Loss: A Common Experience Weight loss is often among the most discussed synthroid side effects due to the fact that nearly everyone wants to look good and maintain a healthy physique. Additionally, many individuals had gained a significant amount of weight prior to using synthroid (as a result of hypothyroidism), a condition which slows metabolism. When metabolic processes are corrected as a result of synthroid, users end up losing weight and the weight loss comes as a pleasant surprise. The primary reason synthroid users experience weight loss is related to an increase in basal metabolic rate (BMR). As was already discussed, individuals with hypothyroidism tend to exhibit abnormally slow basal metabolic rates. When synthroid is taken, basal metabolic rate is corrected (or normalized) and weight loss occurs without much effort. This may result in weight loss without any change in lifestyle, dietary intake, etc. \\
\addlinespace
Analysis & SumRank preserves the \textbf{mixed and more ranking-critical evidence}: it mentions the common belief that levothyroxine may lead to weight loss, but also retains the scientific detail that weight loss is unlikely and that older adults may even gain weight. Qwen2.5-7B \textbf{neglects the important evidence}, which may make the summary \textbf{less reliable} for a query asking ``weight loss or weight gain''. \\
\midrule

\rowcolor{gray!15}
Query & \textbf{Can older adults gain strength by training once per week} \\
\midrule
SumRank(7B) & For beginning exercisers, there is research to support recommending strength training three times, two times, or even once per week. \textbf{The results of this study indicate that training only one or two days per week is effective for increasing muscle strength. However, because the one-day-per-week exercisers endured such a demanding workout (12 sets of bench presses), this training protocol was not very practical.}Consequently, this exercise frequency was ruled out as a sensible strength-training procedure, and people were not advised to work out only once per week. \\
\addlinespace
Qwen2.5-7B & Older adults can gain strength by training once per week, as evidenced by Study three, which showed that one-day-per-week training produced significant strength gains, increasing 15.5 pounds, which was 73 percent as much strength development as the three-day-per-week group (21.2 pounds). However, it is noted that once-a-week trainees must make every effort to avoid missing their workouts, as it is unlikely that an every-other-week exercise session will produce progressive strength development. \\
\addlinespace
Analysis & Both summaries identify that once-per-week training can increase strength, but SumRank better \textbf{preserves the practical evidence}: the once-per-week program was demanding and not recommended. This difference is important for ranking because the query asks whether older adults can gain strength, and a highly relevant document should also \textbf{indicate the condition and limitation of the evidence}. \\
\midrule

\rowcolor{gray!15}
Query & \textbf{how to check if someone has a restraining order} \\
\midrule
SumRank(7B) & No relevant information found. \\
\addlinespace
Qwen2.5-7B & Getting a restraining order against a neighbor can force them to move out, however this depends on the distance between you and your neighbors house. Restraining orders can protect you from your neighbor at a certain distance. If the distance between you and your neighbor is within the distance specified on the restraining order, then he or she will have to move until the order is lifted. \\
\addlinespace
Analysis & SumRank correctly rejects the document because the document discusses the effect of obtaining a restraining order against a neighbor, not how to check whether someone already has a restraining order. Qwen2.5-7B produces a fluent topical summary, but \textbf{the content does not answer the query}. This illustrates that SumRank’s \textbf{rejection behavior can filter topically related but non-relevant evidence}. \\
\midrule

\rowcolor{gray!15}
Query & \textbf{causes of military suicide} \\
\midrule
SumRank(7B) & The suicide rate among recent veterans was 50\% higher than for nonmilitary civilians, a new detailed analysis has found. The rate was slightly higher among veterans who never deployed to Afghanistan or Iraq, suggesting that the causes extend beyond the trauma of war. \textbf{Being white, unmarried, and male were also risk factors.} Men accounted for 83\% of the veterans in the study and all but 124 of the suicides. They were three times more likely than women to take their own lives. Female veterans, however, killed themselves at more than twice the rate of other women. A likely explanation is that women with military experience are much more likely than other women to attempt suicide with firearms, dramatically increasing the likelihood of death. The elevated rate today could reflect differences in who served, with the long wars in Afghanistan and Iraq may have attracted more volunteers prone to risk-taking and impulsive behaviors. Another possibility is that a weak economy during the recent wars made the transition to civilian life more difficult. Experts have also suggested that the military may have become a less forgiving and nurturing place over the course of the wars. The stresses are not limited to the individuals who are sent to war. \\
\addlinespace
Qwen2.5-7B & No relevant information found. \\
\addlinespace
Analysis & SumRank \textbf{preserves multiple pieces of causal and risk-factor evidence.} Qwen2.5-7B incorrectly rejects the document despite its clear relevance. This example shows that SumRank can still generate detailed summaries when the document contains useful relevance evidence. \\
\midrule

\rowcolor{gray!15}
Query & \textbf{what is chronometer who invented it} \\
\midrule
SumRank(7B) & \textbf{The chronometer is an instrument that accurately measures time.} The first one invented was a marine chronometer, and it was created out of the necessity to have an accurate timekeeper for sea travel. John Harrison, a British clockmaker, invented this device in 1737. \\
\addlinespace
Qwen2.5-7B & John Harrison, a British clockmaker, invented the first marine chronometer in 1737. \\
\addlinespace
Analysis & Both summaries capture the key answer, but SumRank \textbf{additionally preserves the definition} of a chronometer and the historical motivation for sea travel. This provides \textbf{more complete evidence for the query}, which asks both what a chronometer is and who invented it. \\

\end{longtable}

\clearpage
\begin{longtable}{p{\dimexpr\textwidth-2\tabcolsep\relax}}
\caption{An example of query-document pair and generated summary from the TREC DL dataset.}
\label{tab:sumrank_output_example_single} \\
\toprule
\endfirsthead

\multicolumn{1}{c}{\textbf{Table \thetable{} -- continued from previous page}} \\
\toprule
\endhead

\bottomrule
\endfoot

\bottomrule
\endlastfoot

\rowcolor{gray!15}

\textbf{Example from TREC DL dataset} \\
\midrule

\textbf{Query:} \\
How do you clean smoke off walls? \\
\addlinespace
\textbf{Document:} \\ How to Clean Smoke Damaged Walls and Ceilings-Water Mitigation Advice \\
\addlinespace
Smoke sponge cleaning smoke off wall \\
\addlinespace
How to Clean Smoke Damaged Walls and Ceilings \\
\addlinespace
Post author: Larry \\
\addlinespace
Post published: 10/10/2019 \\
\addlinespace
Post category: Smoke Issues \\
\addlinespace
Post comments: \\
\addlinespace
0 Comments \\
\addlinespace
How to Clean Smoke Damaged Walls and Ceilings \\
\addlinespace
How to clean smoke damaged walls and ceilings the right way. Yes, there is a correct way to clean smoke damage. Because you can cause more damage, when you don’t understand the does and don’ts of cleaning smoke damage. Therefore, I highly recommend you read ALL my post on the subject or visit my YouTube channel (Link at the bottom). Although, it can be hard work, it is a lot easier and less expensive when you do it the right way the first time. \\
\addlinespace
How to Clean Smoke Damaged Walls and Ceilings \\
\addlinespace
Beware of false information \\
\addlinespace
Because there is a shocking amount of bad information on the internet about how to clean smoke damaged walls. I want to help you understand how to do the job. I have seen a suggestion for using paint thinner. PAINT THINNER! If you use paint thinner on latex walls, you might as well just start sealing and repainting now. Before we can discuss how to clean smoke damaged walls and ceilings, it’s vital to understand the surface you need to clean. Using just water on the wrong surface can cause horrific results. \\
\addlinespace
What will effect how to clean? \\
\addlinespace
Because, some wall surfaces are easier to clean than others, you may not even need to repaint. It depends on how porous the surface of the wall. Also what type of paint and what type of smoke is on the wall. Another problem, has it been cleaned incorrectly before? Smoke damage caused by grease from cooking is especially harder to clean on flat painted walls.? Because many factors can change the cleaning process that you will need to follow it is important to learn how. Certainly, this is not your average cleaning job. \\
\addlinespace
How to Clean Smoke Damaged Walls and Ceilings \\
\addlinespace
Walls, their paint and texture \\
\addlinespace
Typically, the paint on walls and ceilings is latex based. However, some walls use oil-based paint. Other walls may have wallpaper or wood paneling. After all, painted walls have either a flat, semi-gloss or gloss finish. Wood paneling can be sealed or unsealed. Because popcorn ceilings are a whole different ball game when it comes to cleaning. I discuss the methods working with popcorn ceilings in a separate post. 

How to Clean Smoke Damaged Walls and Ceilings. \\
\addlinespace
This article will discuss what materials you need, and how to keep yourself safe. Because I care, please wear proper safety gear while cleaning something as toxic as smoke soot. I am including links at bottom of this post for items you may need. If you have serious smoke damage to your home, we don’t recommend cleaning it yourself. Unless you are not insured. Normally, homeowner’s insurance covers smoke damage. Read more on the basics of smoke damage and working with your insurance company. \\
\addlinespace
Safety First \\
\addlinespace
Certainly, breathing in smoke soot can be toxic to your health. In addition, the chemicals used for cleaning are strong. They can burn skin, eyes, and even lungs if breathed in. On the other hand, the dry cleaning sponge method is the safest. Unfortunately, it does not work on every surface. \\
\addlinespace
When you will be working with chemicals, ensure you have appropriate ventilation. This will require opening windows. Do not run the HVAC when you have smoke damage, this may cause the soot to spread and may damage your system. \\
\addlinespace
What type of surface are your walls or ceiling? \\
\addlinespace
After you determine the type of surface and finish you will be working with, move to the next section of this guide for the cleaning procedures. \\
\addlinespace
Painted walls and ceilings \\
\addlinespace
Before you start, most important thing is to know what type of surface you are working with. You must determine if your painted wall or ceiling uses latex or oil-based paint. Basically, are the walls flat paint, semi-gloss or gloss. Test a tiny area with rubbing alcohol, in an inconspicuous spot. Apply enough rubbing alcohol on the cotton ball so the surface is wet. Rub the alcohol-soaked cotton ball on the wall. If the alcohol removes some of the paint, then your walls use latex paint. No paint on the cotton ball means oil-based paint covers your walls. \\
\addlinespace
If you have oil based painted walls, you can follow the procedure for semi-gloss and gloss painted walls. This is because you can often wash oil based painted surfaces. If your paint is old and wearing, you may want to try the flat based paint cleaning procedure first (Dry smoke sponge). \\
\addlinespace
To be sure of your cleaning method, next determine if the finish is flat, semi-gloss or gloss. Basically, if the paint looks dull, with no shine, than it is a flat finish. Ceilings normally have a flat finish. If there is a any shine to the wall then it has a semi or gloss finish. In that case we treat semi-gloss and gloss painted walls the same when cleaning smoke damage. \\
\addlinespace
Wallpapered walls \\
\addlinespace
While cleaning is possible for many types of wallpaper. However, as wallpaper comes in an extreme variety of textures and porousness. It is impossible to provide accurate information without inspecting the wallpaper in person. \\
\addlinespace
You can try to follow the instructions below, but we caution this, as following the wrong procedure for your surface can cause further damage. If you still want to try at your risk. Because of the risk I would start with the dry smoke sponge, rubbing lightly. Make sure you don’t get the sponge wet, even sweat from your hands can cause smoke to stain. \\
\addlinespace
We recommend either removing the wallpaper or contacting a local specialist to help determine the appropriate cleaning procedure. \\
\addlinespace
Wood paneled walls \\
\addlinespace
The most important factor when dealing with wood walls is if they are sealed or unsealed. \\
\addlinespace
Because sealed wood is easier to try and clean, as the smoke is less likely to embed into the grain of the wood. \\
\addlinespace
You can usually tell unsealed wood by its look and feel. Is it rough? Is it dull? Does it look like it is unfinished? \\
\addlinespace
We do not recommend trying to clean unsealed wood yourself. There is a high risk of permanently discoloring and damaging the wood. It is best to hire a professional experienced in cleaning smoke damage out of unsealed wood. \\
\addlinespace
For sealed wood, you can follow the procedure for cleaning semi-gloss or gloss finished surfaces. However, do not over wet the rag. \\
\addlinespace
Cleaning smoke damaged walls and ceilings – flat paint \\
\addlinespace
Certainly don’t try to wash soot off flat painted surfaces. In the event that the wall is already stained and the sponge is not working. Then and only then try a slightly wet solution. Water or liquid cleaners can instantly turn smoke and soot to black stains, making it even more challenging to clean. Before starting, set up your workspace. \\
\addlinespace
First, Vacuum the floor using a filtered vacuum to clean up any loose soot. Because you do not want to walk and track the soot into more areas. Once vacuumed, Remove furniture and put down drop clothes in the area you are working. Then cover furniture with the light plastic drop clothes. Don’t forget, Put on safety gear, like your respirator mask, this is a must. Fortunately, if you are working with the dry sponge you won’t need to work with any harsh chemicals now. \\
\addlinespace
When you see specks of soot particles on the wall or ceiling, you first gently vacuum them up. Before starting dry sponge cleaning. Hover the vacuum hose over the particles without touching the wall. Because, if you touch the particles, they will likely break and stain the wall. \\
\addlinespace
Clean smoke damaged flat painted walls and ceilings with a smoke sponge \\
\addlinespace
To clean smoke damage off walls and ceilings that have flat paint, you need a dry soot sponge, sometimes called a chemical sponge. Soot sponges are an important tool for soot cleanup and heavily used by cleanup experts. \\
\addlinespace
These sponges feel and work like a large eraser. They are rubbery in texture, used dry, and not like any other type of sponge. Do not try to substitute them for a different product. The sponges are really easy to use, just wipe the area with them; mostly one way. however, DO NOT GET THEM WET. This will ruin the sponge and defeat the purpose. It is important to realize, your hands can’t be sweaty or wet. Make sure your hands are completely dry. Wearing gloves will help with this. Any moisture when cleaning smoke on a flat paint surface will cause staining. \\
\addlinespace
When your sponge gets dirty you can remove a thin layer of the dirty area with rough sandpaper or a fine grater. My favorite, is a piece of rough commercial scrubbing machine pad. If you have a janitorial store near you, buy a rough scrubbing pad used for scrubbing machines. Cut a small piece and rub on the sponge keeping it even. This will keep your sponge in perfect working order. Again, just don’t try to wash it or get it wet. \\
\addlinespace
Usually, painting is not necessary after using the soot sponge. \\
\addlinespace
What if moisture has already touched the flat painted wall or ceiling? When moisture has already touched the wall or ceiling, there will likely be inky looking stains. The smoke sponge is unlikely to remove set stains like these. Go ahead and clean what you can with the smoke sponge before addressing the inky stains. You will need to apply paint sealer to the stain spots on the wall or ceiling after you remove what you can with the smoke sponge. \\
\addlinespace
A really good product to cover smoke stains and odor is Kilz, this is what I use (see below). The product comes in water and oil-based formulas. It does not matter which one you use, but I recommend the water-based version as it covers well and does not have a strong odor. \\
\addlinespace
The stained spots are the only areas you need to seal. It will look better if you feather the outer edges of the sealed spots. Feathering involves lightly brushing in the sealer around the edges, so it blends in the wall. However, stains covering most of a wall or ceiling require sealing the entire wall or ceiling. After the sealer is dry, you can paint the wall. You must paint the whole wall to ensure it blends. \\
\addlinespace
Cleaning smoke damaged walls – semi-gloss or gloss paint \\
\addlinespace
You can also follow this procedure if you are working with sealed wood panels, or oil based painted walls. Doors, door frames, and baseboards usually have a semi-gloss finish, so this procedure should also work for them. \\
\addlinespace
When to clean smoke damage using a liquid solution \\
\addlinespace
To clean walls and ceilings painted with semi-gloss or gloss paint, (surfaces with a slight shine), you will need a liquid cleaning agent made for removing smoke. Prorestore unsmoke degrease all (see below) . If you don’t have grease use the the milder solution of the same product. Minor smoke on semi-gloss or gloss surfaces may clean with household cleaners. \\
\addlinespace
Mix the cleaning agent according to the instructions on the bottle. Use a good lint free cloth that won’t leave fabric on your walls. Test the product in a small inconspicuous area. There should be no discoloration and the paint should not come off or bubble. \\
\addlinespace
Your cloth should be damp but not dripping. Keep rinsing and ringing out your cloth as you work. Do not over wet the surface. Wipe the surface dry with a lint free cloth right after washing the surface to minimize the likelihood of damage. \\
\addlinespace
Grease fire smoke damage \\
\addlinespace
When you have grease built up from grease fire or just a grease build up. Before you try strong solutions on Microwave, stove, counter tops, or cabinets. Try WD40 first then wash that with Dawn dish soap. Reason for this, strong chemicals may cause discolorations. To help with the clean and shinny look, clean last with a little vinegar on the rag then wipe off. \\
\addlinespace
Apply Paint Sealer \\
\addlinespace
Next, apply a paint sealer on any remaining stains. Feather the outer edges of the sealed spots. Feathering involves lightly brushing in the sealer around the edges, so it blends in the wall. \\
\addlinespace
I like Kilz sealers. Kilz comes in water and oil-based formulas. It does not matter which one you use. I recommend the water-based version listed below, Because it covers well and does not have a strong odor. Although you don’t have to seal the whole wall for small spots. Stains covering most of the wall or ceiling may require sealing the entire surface. Be sure to let the sealer completely dry. After the sealer is dry, you can paint the wall. Unlike the sealer you must paint the whole wall to ensure it blends. \\
\addlinespace
In closing \\
\addlinespace
I hope you find this post valuable and were able to salvage your smoke damaged walls and ceiling. I would appreciate a good comment, to help grow my free service. \\
\addlinespace
Visit my YouTube channel \\
\addlinespace
Popcorn Ceilings \\
\addlinespace
Carpet Flooring Upholstery Affiliate disclosure: Watermitigationadvice.com, post may contain affiliate links to help cover cost of operating this website. Thanks, because of your support you are helping to keep the website FREE . For more details about Affiliates or add programs Click here. \\
\addlinespace
Visit our store \\
\addlinespace
Tools and supplies you may need for your job \\
\addlinespace
Face Mask with respirator – Look for a mask rated for smoke filtering with a respirator attached. Make sure the mask fits firmly around your mouth and nose. Do not use a cheap painters’ or dust mask. \\
\addlinespace
Eye Goggles – Sometimes you can find a good mask that will also protect your eyes. If not, you will want a sealed pair of goggles. \\
\addlinespace
Chemical Gloves – Rated for working with heavy-duty cleaning products, these gloves should cover your hands and forearms. \\
\addlinespace
Hooded Coveralls (optional) – Depending on how much smoke damage you are dealing with or if you are working on the ceiling, hooded coveralls can protect you from dripping chemicals. \\
\addlinespace
Filtered Vacuum Cleaner – A vacuum with a good filter will help pick up loose soot without spreading. Without the filter, the vacuum will blow out the smoke soot. Throw away the bag once finished and wash the filter. \\
\addlinespace
Sturdy Ladder – If the area you are working on is out of reach, you will need a sturdy ladder. 4x12 9x12 Light plastic , throw over furniture \\
\addlinespace
Drop Cloths – Less of a safety issue, but drop cloths can help contain the mess. Making sure soot or chemicals do not cause additional damage to floors or furniture. \\
\addlinespace
Kilz, used to cover smoke and water stains, also odor control.Heavy duty smoke and degreaser \\
\addlinespace
For more products visit our store \\
\addlinespace
Tags: Ceiling, Cleaning, Walls \\
\addlinespace
Read more articles \\
\addlinespace
Previous Post How do you clean wet carpet? \\
\addlinespace
Next Post Cleaning Smoke Damaged Upholstery \\
\midrule
\rowcolor{gray!15}
\textbf{Model Output by SumRank} \\
\midrule
Smoke sponge cleaning smoke off wall
\newline\newline
To clean smoke damage off walls and ceilings that have flat paint, you need a dry soot sponge, sometimes called a chemical sponge. Soot sponges are an important tool for soot cleanup and heavily used by cleanup experts. These sponges feel and work like a large eraser. They are rubbery in texture, used dry, and not like any other type of sponge. Do not try to substitute them for a different product. The sponges are really easy to use, just wipe the area with them; mostly one way. However, DO NOT GET THEM WET. This will ruin the sponge and defeat the purpose. It is important to realize, your hands can’t be sweaty or wet. Make sure your hands are completely dry. Wearing gloves will help with this. Any moisture when cleaning smoke on a flat paint surface will cause staining. When your sponge gets dirty you can remove a thin layer of the dirty area with rough sandpaper or a fine grater. My favorite, is a piece of rough commercial scrubbing machine pad. If you have a janitorial store near you, buy a rough scrubbing pad used for scrubbing machines. Cut a small piece and rub on the sponge keeping it even. This will keep your sponge in perfect working order. Again, just don’t try to wash it or get it wet. Usually, painting is not necessary after using the soot sponge. \\
\end{longtable}

\clearpage

\begin{longtable}{p{\dimexpr\textwidth-2\tabcolsep\relax}}
\caption{An example of Listwise Rerank of Summaries from the TREC DL dataset.}
\label{tab:listwise_output_example} \\
\toprule
\endfirsthead

\multicolumn{1}{c}{\textbf{Table \thetable{} -- continued from previous page}} \\
\toprule
\endhead

\bottomrule
\endfoot

\bottomrule
\endlastfoot

\rowcolor{gray!15}
\textbf{Example from TREC DL dataset} \\
\midrule

 \textbf{Query:} \\
        How do you clean smoke off walls? \\

\textbf{Documents:} \\

[1] Smoke sponge cleaning smoke off wall
        
To clean smoke damage off walls and ceilings that have flat paint, you need a dry soot sponge, sometimes called a chemical sponge. Soot sponges are an important tool for soot cleanup and heavily used by cleanup experts. These sponges feel and work like a large eraser. They are rubbery in texture, used dry, and not like any other type of sponge. Do not try to substitute them for a different product. The sponges are really easy to use, just wipe the area with them; mostly one way. However, DO NOT GET THEM WET. This will ruin the sponge and defeat the purpose. It is important to realize, your hands can’t be sweaty or wet. Make sure your hands are completely dry. Wearing gloves will help with this. Any moisture when cleaning smoke on a flat paint surface will cause staining. When your sponge gets dirty you can remove a thin layer of the dirty area with rough sandpaper or a fine grater. My favorite, is a piece of rough commercial scrubbing machine pad. If you have a janitorial store near you, buy a rough scrubbing pad used for scrubbing machines. Cut a small piece and rub on the sponge keeping it even. This will keep your sponge in perfect working order. Again, just don’t try to wash it or get it wet. Usually, painting is not necessary after using the soot sponge.\\
\\

[2] 6 Quick Ways to Clean Nicotine Off Walls

1. Cleaning Cigarette Smoke from Walls with Vinegar

Using pure vinegar is an excellent way to cut through the worst nicotine residue stains, as the acetic acid breaks down the stain, plus it helps in removing the stale smell. There are numerous ways you can use vinegar to clean nicotine off your walls. Add a few drops of your favorite essential oil to the mixture if the vinegar smell is unpleasant for you.

2. Remove Smoke Residue from Walls with a Magic Eraser

With a Mr. Clean Magic Eraser, all you need is the sponge and some water to remove crayon from walls or to eliminate nicotine or other stains. The scrubbers on these sponges are activated by the water and power through any cigarette stain on your drywall, but also clean glass, doors, baseboards, and more.

3. Clean Nicotine Off Walls using Trisodium Phosphate

Considered an extreme solution, Trisodium Phosphate tackles the worst nicotine stains, as it is an effective degreaser for cigarette tar. Trisodium phosphate, also called TSP, often works where other cleaning solutions don’t as long as it’s used correctly.

4. Chomp Painted Wall Cleaner Removes Heavy Smoke Stains

Nobody wants to repaint their walls if they don’t have to. When it comes to dealing with the after-effects of a heavy smoker, you often think your only option is repainting. Before you opt for this drastic solution, try one of the newest cleaning products called Painted Wall Cleaner by Chomp.

5. Clean up Nicotine with Baking Soda

Before you clean walls, use a vacuum with the upholstery attachment to clean all the loose dirt and debris. You may also use a damp chamois cloth to wipe down the dust or a microfiber cloth.

6. Cleaning Nicotine Stains with Simple Green

Safe to use on all painted walls, Simple Green has proven effective on all kinds of smoke damage, including years of smoking inside the home. Simple Green powers through those yellow stains and works as a deodorizer, helping to eliminate the old cigarette smell.\\
\\

[3] No relevant information found.\\
\\

[4] No relevant information found.\\
\\

[5] I use Clorox cleaner with bleach. 

I'm currently cleaning an estate, and this is PROOF that nicotine can be washed off of the walls with vinegar, and ammonia. I didn't even use the baking soda with the concoction! Also, if you use a Scotch-Brite Dobie, it works much better than just using a rag!

Half vinegar half water will not only cut the nicotine from the walls while leaving the paint on, it will also remove the odor of the nicotine. Yes, the vinegar smell will dissipate overnight. Promise.

I use a cleaner called Awesome, sold in Dollar General, Family Dollar, etc. stores. Very cheap yet very good.

Use a little ammonia, a little white vinegar, a squirt of dawn dish soap, 1/4 cup of baking soda, and some hot water - at least one gallon of hot water. This takes nicotine off your walls fast and you don't have to go back over it to rinse. I have even used Easy Off Oven Cleaner in the blue can on stubborn areas on the wall. With that though, you have to work very fast because it will remove the paint right off of the wall.

I used white rubbing alcohol and a handi-wipe. Because alcohol evaporates quickly that was the only cloth that worked. I completely soaked the rag no need to wring out and wiped down wall. Follow immediately with a wet towel (not one of the good one's) . The nicotine came right off. I still have the handiwipe but the towel had to be tossed. Good Luck.

Half water and have white distilled vinegar applied freely and repeatedly until the sponge or towel shows no nicotine residue - just wait till you see how well this cuts and removes the truly icky yellow mess! The vinegar 'aroma' will dissipate in about a half hour after the last wipe-down leaving the room/house odor free.

1 Cup Ammonia 1/2 Cup Vinegar 1/4 Cup Baking Soda 1 Gallon Hot Water

I mix the ammonia, vinegar, water first then slowly add the baking soda. Do this over the sink as it can fizz up. Using a large sponge or rag, first wipe with the solution then wipe again with clear water. I only wipe it again if it is a real heavy build up your trying to remove.\\
\\

[6] Steps to Clean Cigarette Smoke on Walls

1. Choose Your Cleaning Method

When choosing a wall cleaning method for removing cigarette smoke, there are several different solutions you can try, including:

White vinegar

Ammonia

Trisodium phosphate (TSP)

Mr. Clean Magic Eraser

Various other wall cleaning products

Do some research into the right solution for your home. Remember that stronger chemicals like ammonia and TSP will work wonders on large, tough stains — but they are strong, so use them with caution (and protection!).

If you’re cleaning a small or lightly soiled area, you can probably get away with a gentler solution, like the Magic Eraser.

2. Assemble All Supplies

Once you’ve chosen your preferred wall cleaning method, get all your materials together. Here’s a list of some good items to have on hand when removing smoke from walls:

Cleaning solution

Spray bottle

Large bucket

Sponge mop

Scrub broom

Drop cloth

Rags Gloves Primer Paint 

If your walls have latex or oil-based paint, stay away from scrub sponges, harsh chemicals, or Teflon pads, which can all leave scratch marks on your walls.

3. Prep the Area

Got everything? Good. Now you can prep the area for wall cleaning.

Start by laying down your drop cloth in case you drip solution on the floor while cleaning. Be sure to cover the baseboards particularly well. Then, switch off the power in the room just to be safe. You’ll want to be sure you don’t get water or solution near electrical outlets while you’re working. Finally, dust the walls by wiping them with a microfiber or damp chamois cloth.

4. Wash and Dry the Walls

Now you’re ready to get to work! (We hope you’ve had your coffee, because this could take a while. . . .) The process of cleaning smoke from walls will differ depending on the type of cleaning agent you choose. Here are the five general steps to follow: Soak a smooth towel or rag in the solution for a few seconds. Starting at the bottom of the wall, vigorously scrub a small section (about four square feet) in circular motions, working your way up the wall as you scrub. Rinse the section with a clean cloth dampened with water. 

Immediately after rinsing, dry the section with a soft, clean cloth, wiping up any drips.\\
\\

[7] Steps to clean smoke off walls:

1. Ensure the area is properly ventilated.

2. Wear safety goggles and heavy-duty gloves.

3. Place a drop-cloth on the floor.

4. Remove loose soot with a vacuum cleaner or a dry chemical sponge.

5. Use Trisodium Phosphate (TSP) solution: Mix 1 tablespoon of TSP with 1 gallon of warm water. Wet a sponge with the solution and wipe the wall vigorously. Rinse with a clean rag.

6. Consider using commercial cleaning products designed for smoke damage.

7. Attempt to remove the smoke odor with household items like white vinegar or Febreeze.\\
\\

[8] How To Clean Smoke Stained Walls Before Painting

I’ve been in many homes with smoke stained walls either from smokers, fireplaces or just burnt cooking in the kitchen.

Smoke needs to be cleaned up and properly sealed before painting if you want to completely eliminate the smell and prevent the smoke from bleeding through your paint into your new paint.

While there are many different thoughts on how to best clean smoke, I’ve had the best luck with Arm and Hammer’s Cleaning Soda.

Cleaning soda (sodium carbonate) will help clean and remove the smoke and the smoke smell.

Simply mix the cleaning soda with warm water and scrub down the walls with a rag.

To rinse the walls, mix a cup a vinegar with a gallon of warm water and wash the walls again. This will neutralize the washing soda and help kill the smoke smell (and disinfect the walls).

Once the walls dry, sand the walls using a pole and disc sander as mentioned above. Wipe off the dust and you’re ready to paint.

NOTE: If your walls have really bad smoke stains, you may need to prime the walls to seal the smoke stain and prevent bleedthrough. I recommend Killz Max for this.\\
\\

[9] Wash or Clean All Clothes, Fabrics, and Linens

Clean Your Ducts and Replace Filters (Air, Furnace, and Air Conditioning)

Clean Your HVAC Evaporator Coil

Wash or Clean All Furniture and Surfaces

Wash Your Walls Down

Clean the Baseboards and Various Fixtures

Prime and Paint Your Walls

Clean or Replace the Carpet

Leave Out Some Deodorizers

For washing walls, use a TSP solution: 1. In one 5-gallon bucket, mix TSP with water according to the package instructions. 2. In the other bucket, fill with plain water. 3. Place large rags underneath the section of the wall to protect the baseboards. 4. Soak a scrub brush in the TSP mixture, scrub the walls aggressively from bottom to top, then rinse with the clear water from the other bucket, wiping from top to bottom. Change the water often as it becomes dirty.\\
\\

[10] No relevant information found.\\
\\

[11] Using vinegar can be an excellent cleaning agent to treat smoke damaged walls. Here’s how to use it: 1. Warm the vinegar and add it to an equal volume of warm water. If the yellowing is particularly harsh, increase the ratio of vinegar to water. 2. Pour the diluted vinegar solution into a spray bottle. 3. Spray the smoke stain with the diluted vinegar solution. 4. Walk away, drink a cup of coffee, and relax for a few minutes. This will give the diluted vinegar solution time to work. 5. Scrub the area using a clean, damp cloth. 6. Repeat steps 1 to 5 as often as necessary, until your wall is clean of smoke damage.\\
\\

[12] How to Remove Soot from Walls And Clean The Stains

Firstly, it’s important to state that soot buildup usually is a result of smoky fires. When smoke billows into the home, it leaves residue on furniture, floors, ceilings and walls. To avoid excessive buildup, make sure your fireplace is working properly. Most of the smoke generated by a fire should be vented up through the chimney and to the outside. Soot removal needs to be approached with caution. Many people think it’s similar to dusting when in fact soot can be a real physical health hazard to people around it. Soot is, after all, a carbon residue. Anytime you attempt to clean off soot from walls, make sure you’re wearing a face mask and safety glasses or goggles. They’ll stop the soot from getting into your lungs or eyes and triggering asthma or some other health condition. Also, wear old clothes when you’re taking on the soot because you’re going to get dirty. Open some windows or get an air purifier in the room so it will prevent soot particles that are flung off the wall during cleaning from settling elsewhere. You might want to consider removing furniture and other items for the room as well. Inevitably, as you scrub soot off the walls, some of it will become airborne and settle on couches, tables or chairs. In order to protect any carpets or exposed floors, lay down a plastic sheet to catch falling soot particles. You can tape sheet to molding and other flat surfaces to stop buildup as well. 

The Soot Removal Process Strictly follow this method of cleaning to avoid any smear marks on your walls. It’s especially important for removing soot on painted walls, because they’re more delicate and susceptible to markings. First, use a chemical dry sponge to wipe the affected walls from top to bottom. Chemical sponges are also referred to as dry cleaning sponges. It just means they are made from vulcanized rubber. They’re widely available in stores and online. Do not rub in circular motion or use short strokes. Try to use steady, long downward swiping motions. The dry sponge knocks off the soot that’s stuck on the wall that may not yet be visible. Eventually, the surfaces of the sponge will become saturated with soot. You can either switch to a new sponge or use scissors to cut off the top layers and continue using the same sponge.\\
\\

[13] For caked-on smoke from grease fires or smokers in the house, mix 1/4 cup of baking soda, 1 cup of ammonia and 1/2 cup of vinegar in a gallon bucket filled with the hottest water you can stand. Wearing rubber gloves dip a sponge's abrasive side into the mixture and wash the walls from the top down; this helps you to clean up drips or runs as you go. Keep a second bucket full of hot rinse water so you can rinse with a clean sponge. Do not let drip or run marks stay on the enameled walls.

Anyone who uses a fireplace can forget to open the damper, causing the smoke to fill the home or room. Wood smoke is not as hard to remove as cigarette smoke, but it still may contain wood resins and tar. Mix 1 cup of clear ammonia in a gallon bucket half-full of warm water. Wash the walls from the top down, starting in the left-hand corner farthest from where you exit the room and work toward the exit. Rinse with a soft cloth dipped in warm water. If you are sensitive to the fumes from the ammonia, wear a face mask as you clean.

A steam-cleaning machine intended for use inside the house can make cleaning smoke from fire damage a lot easier. The machine boils water to the point of steam, which exists through a tube with a cleaning applicator on the end of it. Once steam starts coming out of the machine, run the applicator down the wall. Frequently change out the cleaning cloth on the applicator to keep from spreading the soot and smoke back onto the wall. A steam cleaner can also help to loosen caked-on smoke or its sticky residue. Scrape off the sticky residue from the walls into a bucket. Rinse clean with a dampened sponge or cloth.

Though some states regulate or prohibit the use of trisodium phosphate, new versions make it safer to use if you can buy it in your state. But wear gloves and a face mask for protection when cleaning with TSP, because it is a powerful chemical cleaner. To clean furnace smoke buildup, dilute one heaping teaspoon of TSP in a gallon of warm water, mixing it thoroughly until it dissolves. Wipe the walls from the top down with a sponge wrung out after you dip it in the solution. Follow by rinsing with a clean sponge wrung out in clean water.\\
\\

[14] Vinegar: Odor Eater

Regular white vinegar removes odors and cleans the ceiling and walls a bit in the process. Dip a sponge or cloth into a bowl of vinegar, wringing out most of the liquid; then wipe down the entire surface of each wall and ceiling in the room. Open the windows or turn on a ceiling fan to hasten the drying process, if desired. If the odor persists, give the room a second vinegar wipe-down treatment.

Heavy-Duty Cleaner

Smoke residue clings to walls and ceilings after a grease fire, meaning every inch of those surfaces requires cleaning. If the walls or ceiling feel greasy or sooty, a treatment with TSP (tri-sodium phosphate) gets them clean once again. Mix a tablespoon of TSP into a gallon of warm water in a bucket and drop a large sponge into the water. Another bucket full of clean water with its own sponge is necessary as well, as the TSP has to be wiped back off the surfaces quickly to prevent paint discoloration. While wearing rubber gloves and eye protection, wring most of the liquid out of the sponge and scrub the surfaces, cleaning small areas at a time; then wipe down the same areas using the water-dampened sponge as a rinse. Keep the sponges only as moist as necessary, as soaking the walls or ceiling may damage them.

A Chemical Sponge Solution
A special type of sponge is designed specifically to remove soot and stains related to smoke damage after a fire. While it’s called a sponge, it works more like an eraser when you rub it over dry walls to remove marks. Rub down the ceiling, overlapping strokes slightly with the chemical sponge until the entire ceiling has been treated; then work on the walls, cleaning from the top down, as the sponge may drop a bit of debris in the process. When the sponge appears dirty, slice the dirty areas away with a utility knife to reveal a fresh surface.\\
\\

[15] Soot stains will come off with the right cleaner. The key is to use cleaners that are strong enough to remove the smoke stains without peeling or scrubbing off the paint. Step 1: Rub a dry cleaning sponge on the walls, starting at the top and working your way down. Do not wet the sponge; simply rub it on the walls. Wear rubber gloves as you work. Step 2: Cut off a layer of the dry cleaning sponge as it becomes clogged with soot. Use a sharp razor to cut the top layer off and to reveal a fresh layer. Continue working. Step 3: Wash the walls with a household cleaner or cleaner specially designed to remove smoke odors. Fill a bucket with the cleaner after mixing it with water according to the manufacturer's instructions. Wash 1-foot square sections of the wall at a time using a regular sponge. Step 4: Rinse the wall with plain water and a clean sponge to remove any cleaner residue. Let it air-dry thoroughly. Vacuum the wall first if you have loose soot. A shellac primer used before painting covers both smoke stains and odors.\\
\\

[16] 3 Steps To Clean Soot Off Walls

1- Clean The Soot 

Removing black soot on walls can be quite challenging particularly if it’s a bbq smoke stain or fireplace soot. To clean soot off your walls, you will need to have a good soot remover sponge along with water and mild detergent or you can use TSP, but only if it is well-diluted. Remember that drywall is susceptible to water, so you only want to dampen the sponge. Start wiping the walls and surfaces that are covered in soot. Keep in mind that your best efforts will probably not get rid of all the soot that has been building up over the years. This is because the soot has become embedded in the paint itself and unless you want to replace the drywall, there will still be a stain. However, your efforts with the sponge will remove soot that has been newly placed and will clean the wall sufficiently to start the covering process.

2- Primer The Wall Once the black smoke from the walls has been fully cleaned and dried, you are now ready to apply the alkyd primer to the surface. It is important that you choose a primer (like KILZ MAX or Zinsser BIN) that offers some stain-covering abilities as noted on the can itself. Such primers for smoke damage are designed to not only cover the stains but lock in any odors that they generate. Use a disposable paintbrush to apply the primer to the areas that have soot stains. After it dries, apply a single coat of primer to the entire surface that you want to repaint. Allow it to dry for a couple of hours, then apply a new coat. You’ll want to apply enough coats so that the primer you used to specifically cover the stains is now fully covered itself. Wait until it all dries before covering with the proper paint.

3- Paint The Wall The paint you choose to remove soot from walls should be acrylic latex paint that can be scrubbed. Be sure to purchase such a smoke resistant paint (on Amazon) because this will not let any new soot stick and embed itself to the surface. Use a good quality roller or brush to apply the paint to avoid any lap marks. You’ll want to apply at least two coats of paint to ensure that the surface is properly sealed. Once the paint is dry, check it over to see if the coats are even and the paint looks right.\\
\\

[17] No relevant information found.\\
\\

[18] No relevant information found.\\
\\

[19] No relevant information found.\\
\\

[20] No relevant information found.\\
\\

\midrule
\rowcolor{gray!15}
\textbf{Model Output} \\
\midrule

[2] > [6] > [7] > [13] > [15] > [14] > [9] > [11] > [5] > [8] > [1] > [12] > [16] > [3] > [4] > [10] > [17] > [18] > [19] > [20]
\end{longtable}

\clearpage             
\twocolumn

\end{document}